\documentclass[11pt]{article}

\usepackage[a4paper, margin=1in]{geometry} % Standardizes page margins

\usepackage[utf8]{inputenc} % Supports UTF-8 encoding
\usepackage[T1]{fontenc}    % Supports proper encoding
\usepackage[english]{babel} % Sets the document language to English

\usepackage{setspace}       % For line spacing
\usepackage{graphicx}       % For including graphics
\usepackage{longtable}      % For long tables
\usepackage{lscape}         % For landscape tables/pages
\usepackage{float}          % For floating objects

\usepackage{amssymb}        % For mathematical symbols
\usepackage{amscd}          % For commutative diagrams
\usepackage{amsfonts}       % For additional math fonts
\usepackage{amsmath}        % For advanced math typesetting
\usepackage{mathrsfs}       % For script math fonts
\usepackage{bbm}            % For blackboard bold math symbols

\usepackage{url}            % For formatting URLs
\usepackage{hyperref}       % For hyperlinks
\hypersetup{
    colorlinks=true,
    linkcolor=blue,
    citecolor=blue,
    filecolor=blue,
    urlcolor=blue
}
\usepackage[numbers]{natbib} % For numerical citations

\usepackage[margin=.5cm,font=footnotesize]{caption} % Custom captions

\usepackage{lineno}         % For line numbering

\usepackage{appendix}       % For appendix management
\usepackage{subfig}         % For sub-figures

\usepackage{tcolorbox}      % For colored boxes

\usepackage{amsthm}         % For theorem environments

\renewcommand{\contentsname}{Table of Contents}

\title{\textbf{Algorithmic Idealism III: "Algorithmic State" Formulation of Quantum Mechanics}}
\author{Krzysztof Sienicki\thanks{Chair of Theoretical Physics of Naturally Intelligent Systems, Lipowa 2/Topolowa 19, 05-807 Podkowa Leśna, Poland, EU.}}
\date{\today}

\begin{document}

\maketitle
% Table of contents
\renewcommand{\contentsname}{Table of Contents}
\tableofcontents

\begin{abstract}
This work introduces Algorithmic Idealism, a framework that reinterprets quantum mechanics as a computational process governed by algorithmic probability, informational simplicity, and utility optimization. Reality is modeled as an agent-environment feedback loop, where agents maximize utility by predicting self-state transitions. Core quantum phenomena - such as measurement, entanglement, and probabilities - are explained through informational constructs. The paper presents a set of quantum mechanics postulates inspired by this framework and explains their mathematical meaning in detail. These postulates include measurement as Bayesian updating, entanglement as joint utility optimization, and quantum probabilities as utility-weighted predictions. Physical laws are shown to emerge as informational regularities from algorithmic constraints. By defining identity through decision-making consistency and treating simulated and base realities as indistinguishable, the framework also resolves philosophical questions about identity and reality. This approach unifies quantum mechanics with computational theory, offering a novel perspective on foundational physics and its connection to information and computation.
\end{abstract}
\section{Introduction}

Quantum mechanics, developed just over a century ago, remains one of the most profound and far-reaching theories in modern science. Since its inception in the early twentieth century, with pivotal contributions from Planck, Einstein, Schrödinger, Heisenberg, and Dirac, quantum mechanics has provided a framework for understanding and predicting the behavior of particles at microscopic scales. Beyond its immense success in physics, quantum mechanics has inspired numerous interpretations and reformulations, reflecting its conceptual richness and foundational depth.

The strength of the theory lies not only in its predictive power but also in its adaptability to various axiomatic structures and philosophical perspectives. Over the decades, physicists and mathematicians have sought alternative formulations of quantum mechanics, each highlighting different principles and motivations. For example, the original Hilbert space formalism \cite{Wigner1967} was complemented by Feynman's path-integral approach \cite{Feynman1948}, while information-theoretic derivations \cite{Chiribella2011}, \cite{Hardy2003} aimed to reframe the theory in terms of informational and causal principles. Interpretative frameworks such as Everett's many-worlds\cite{Everett1957}, Rovelli's relational quantum mechanics\cite{Rovelli1996}, and Zurek's quantum Darwinism \cite{Zurek1989} further underline the conceptual diversity within the quantum landscape.

John Wheeler’s influential “It from Bit” thesis redefined the relationship between information and quantum mechanics, proposing that the universe's foundation lies in informational processes rather than material entities. He famously asserted that, “every it—every particle, every field of force, even the spacetime continuum itself—derives its meaning and existence from answered yes-or-no questions, binary choices, bits” \cite{wheeler1990}. In Wheeler’s view, physical reality is not fundamental but emerges from the interplay of informational interactions.

A cornerstone of Wheeler’s thesis is the active role of the observer. Through measurements, observers pose binary questions, and the responses define the properties of quantum systems. This perspective aligns with quantum mechanics, where measurement plays a central role in determining system states. Wheeler viewed quantum mechanics as fundamentally informational, with principles such as superposition and entanglement illustrating how quantum systems process information in ways unattainable by classical systems \cite{jaeger2019}.

Wheeler also applied his "It from Bit" concept to black holes, arguing that they are informational entities, with their surface area encoding entropy. This view reflects his broader claim that space and time are emergent constructs, derived from quantum informational processes rather than being fundamental components of reality \cite{bekenstein1973}. He proposed that unifying general relativity and quantum mechanics would require an informational framework, where spacetime itself arises from deeper informational principles.

Wheeler’s ideas have profoundly influenced modern physics, inspiring informational reconstructions of quantum mechanics, which seek to derive its formalism from simple informational axioms. His vision has also fueled interpretations of the universe as a computational system or quantum computer, further cementing his thesis's enduring relevance.\cite{Zurek1989}

By reframing the universe as fundamentally informational, Wheeler provided a revolutionary lens through which to understand reality. His thesis not only reshaped foundational physics but continues to inspire research into the informational essence of the universe.

Quantum mechanics has been interpreted and reconstructed using two principal approaches. The interpretational approach seeks to define the meaning of the elements of quantum mechanics and often delves into metaphysical implications. In contrast, the reconstructive approach focuses on mathematically deriving quantum mechanics from simple operational or informational principles, postponing interpretational concerns. Reconstruction emphasizes foundational principles such as superposition and entanglement, which allow quantum systems to encode and manipulate information in ways that differ fundamentally from classical systems. These unique properties, such as nonlocal correlations that violate Bell-type inequalities, are central to the informational perspective on quantum mechanics.\cite{jaeger2019}

Informational reconstructions often invoke three foundational positions: informational ontology, which posits that everything is reducible to information; digital ontology, where the universe is fundamentally discrete and computational; and pancomputationalism, which considers the universe as a computational system analogous to a Turing machine. John Wheeler’s "It from Bit" thesis articulates a profound connection between physical reality and information, asserting that physical entities, including space-time, derive from binary informational processes (bits). This concept underpins numerous efforts to unify quantum mechanics, information theory, and even quantum gravity .\cite{wheeler1990}

The idea that physical reality is deeply intertwined with information, as articulated by John Wheeler's "It from Bit" thesis, offers a profound framework for understanding the connection between quantum mechanics, information theory, and the structure of the universe. Informational reconstructions frequently build upon three key positions: informational ontology, which posits that everything is reducible to information; digital ontology, which views the universe as fundamentally discrete and computational; and pancomputationalism, which treats the universe as a computational system akin to a Turing machine. However, the extension of these ideas into the assertion that "the universe is a computer" represents an overreaching conclusion. \cite{Lloyd2013}, \cite{Weinberg2002},\cite{Wharton2015}, \cite{Schmidhuber2006} While such a claim captures the poetic appeal of computational metaphors, it risks conflating conceptual models with ontological reality. Rather than equating the universe to a literal computer, it is more accurate to interpret these frameworks as metaphors or tools for understanding how information and computation can model and describe physical phenomena.

Wheeler's thesis remains a foundational insight for exploring how binary processes (bits) underpin the structure of physical entities, including space-time itself, but care must be taken to distinguish between modeling frameworks and the nature of reality itself.

Key advances in reconstructing quantum mechanics include the Popescu-Rohrlich framework, which explores fundamental constraints like nonlocality and causality, leading to the concept of "superquantum" correlations that exceed traditional quantum mechanical limits but remain consistent with relativistic causality'\cite{Popescu1994} Similarly, Rovelli’s relational quantum mechanics suggests that information and its relevance depend on the observer, deriving quantum mechanics from logical axioms.\cite{Rovelli1996} The Clifton-Bub-Halvorson (CBH) theorem provides a significant informational foundation for quantum mechanics, deriving it from three key constraints: the impossibility of superluminal signaling, the impossibility of perfect state cloning, and the impossibility of unconditionally secure quantum bit commitment.\cite{Clifton2003}

Reconstruction through axiomatic methods has also yielded significant insights. Lucien Hardy’s framework introduces five axioms to derive quantum theory, emphasizing probabilities and information. These axioms lead to the formal structure of quantum mechanics, including the derivation of Hilbert spaces.\cite{Hardy2003} Similarly, the axiomatic framework proposed by Masanes and Müller minimizes assumptions, using operational principles such as state preparation, reversible transformations, and measurements to reconstruct quantum mechanics as a generalized probability theory.\cite{masanes2011} Both approaches highlight the role of entanglement and operational constraints in shaping the structure of quantum mechanics.

Quantum mechanics' informational perspective reveals unique features, such as entanglement and the purification principle. Entanglement exemplifies quantum holism, where the composite system encodes more information than its parts. The purification principle states that every mixed state arises from discarding parts of a larger pure system, emphasizing the conservation of information. These concepts illustrate the deep interplay between information and the physical world.\cite{Bokulich2010}

Despite its successes, informational reconstruction faces challenges. There is ongoing debate over whether information is more fundamental than physical matter or whether physics merely constrains information. Additionally, many approaches still require interpretational elements, highlighting the unresolved philosophical questions underlying quantum mechanics. For instance, the CBH theorem shows that quantum information must be treated as a physical primitive to connect the informational constraints with the quantum formalism .\cite{Clifton2003}

Informational reconstruction offers promising directions for advancing our understanding of quantum mechanics. Hardy’s framework points toward a potential unification of quantum mechanics and gravity through informational principles. Further exploration of generalized quantum theories, such as real-vector-space quantum mechanics, holds the potential to deepen our understanding of quantum systems and their foundational principles. These efforts could also pave the way for advancements in quantum computing, cryptography, and fundamental physics.\cite{Hardy2003}

This article builds upon this tradition by exploring how Algorithmic Idealism, grounded in principles of algorithmic probability and informational simplicity, provides a unified framework for quantum mechanics. By treating quantum states, measurements, and transitions as informational constructs, Algorithmic Idealism offers a novel perspective on quantum theory's foundational aspects. Specifically, we investigate how algorithmic simplicity, coherence, and Bayesian updating can serve as postulates to derive quantum mechanics and address long-standing questions about its interpretation.

\section{What is Algorithmic Idealism?}

Algorithmic Idealism, recently developed by Markus Müller and the present author \cite{Müller2024}, \cite{Müller2020}, \cite{Sienicki2024_I}, and \cite{Sienicki2024_II}, is a theoretical framework that reconceptualizes reality as an emergent phenomenon arising from informational and computational processes. Rather than treating the universe as an independently existing external construct, Algorithmic Idealism emphasizes subjective experience, computational principles, and algorithmic information theory. It asserts that reality is best understood as a dynamic sequence of transitions between "self-states," which represent an agent's complete informational configuration, encompassing its beliefs, observations, memories, and decision-making processes.

At its core, Algorithmic Idealism draws from algorithmic probability, particularly Solomonoff induction, to explain how agents predict future experiences. Solomonoff induction prioritizes simpler (low-complexity) hypotheses, offering a framework for understanding how agents optimize their predictions by balancing simplicity and accuracy. This leads to a shift in focus from questions about the objective existence of reality to the subjective question: "What will an agent experience next?" Algorithmic Idealism rejects the traditional distinction between simulated and base realities, arguing that from the agent's perspective, informational coherence is all that matters. As long as the transitions between self-states are consistent and meaningful, there is no practical difference between simulation and physical reality.

A crucial aspect of Algorithmic Idealism is its reinterpretation of quantum mechanics. It re-frames quantum phenomena, such as measurement and entanglement, in terms of informational principles and utility maximization. Measurement, for instance, is not seen as a collapse of an objective wavefunction but rather as a Bayesian update to the agent's probabilistic model of reality. Similarly, quantum probabilities are understood as utility-weighted predictions that optimize the agent's expectations about future self-states. Entanglement, a hallmark of quantum mechanics, is interpreted as a shared optimization of utility between subsystems, which explains non-local correlations in a way consistent with the computational principles of the framework.

Another significant contribution of Algorithmic Idealism is its explanation of physical laws. Instead of treating physical laws as fundamental truths about the universe, the framework sees them as emergent regularities that arise from the dynamics of self-state transitions. For example, conservation laws and symmetries in physics can be understood as stable patterns that emerge naturally from the underlying informational processes. This perspective provides a computational foundation for quantum mechanics and bridges the gap between physics and algorithmic information theory.

Algorithmic Idealism also addresses long-standing philosophical challenges. It resolves paradoxes like the Boltzmann Brain problem \cite{Sienicki2024_I} by eliminating the need for self-location in an external universe. Instead, the framework focuses on informational coherence and utility, avoiding issues related to improbable or arbitrary states. Similarly, it offers a fresh perspective on thought experiments such as Parfit's Teletransportation Paradox. \cite{Sienicki2024_I} In this view, identity is not tied to physical continuity but to the informational integrity and consistency of self-states. This redefinition of identity has profound implications for understanding consciousness, cloning, and the ethical treatment of digital beings or simulated entities.

The "ethical" implications of Algorithmic Idealism extend into considerations about "moral" responsibility, simulated realities, and the preservation of informational integrity. By redefining knowledge as predictive utility rather than absolute truth, the framework raises important questions about how agents should navigate a reality that is fundamentally computational and informational. It also challenges traditional ethical paradigms by emphasizing the continuity of self-states as the primary criterion for moral consideration, which has significant implications for emerging technologies such as artificial intelligence and virtual reality.

Practically, Algorithmic Idealism offers a unifying perspective that connects foundational physics, philosophy, and computational theory. It has the potential to address some of the deepest questions about the nature of reality, consciousness, and identity. For instance, the framework explains why physical laws exhibit the regularities we observe, why quantum mechanics is probabilistic, and how subjective experience arises from informational processes. By treating reality as a computationally optimized construct, Algorithmic Idealism provides a rigorous and flexible approach to understanding existence in an increasingly digital and interconnected world.

\section{Postulates of Algorithmic State Quantum Mechanics}

Here is a set of postulates for quantum mechanics inspired by \textbf{Algorithmic Idealism}, a conceptual framework emphasizing that physical laws and phenomena emerge from informational and computational principles. This perspective treats the universe as fundamentally informational and computational in nature, with quantum mechanics as the governing "algorithm."
\bigskip
\subsection{Postulate 1.}
\bigskip
\begin{tcolorbox}[colframe=blue!30, colback=blue!10, coltitle=black, 
title=\textbf{Postulate 1: Reality as Agent-Environment Interaction}]
\textit{Reality consists of an agent interacting with its environment, modeled as an informational feedback loop where the agent predicts future self-states to maximize its expected utility.}
\end{tcolorbox}
\bigskip

\textbf{Context of Postulate 1.} The universe is not an objective external construct, but emerges from the computational model of the agent's environment, where experiences are determined by algorithmic transitions. The universe operates as a computational process, where states and dynamics are governed by fundamental informational algorithms. Physical systems are fundamentally described by discrete or continuous bits of information, and quantum mechanics emerges as the optimal algorithm to encode and evolve this information. 
\bigskip

\subsection{Postulate 2.}
\bigskip
\begin{tcolorbox}[colframe=blue!30, colback=blue!10, coltitle=black, 
title=\textbf{Postulate 2: Self-States as Information Processing Units}]
\textit{A self-state represents the complete informational configuration of the agent, encoding its current beliefs, memories, and observations.}
\end{tcolorbox}
\bigskip

\textbf{Context of Postulate 2.} The state of a quantum system encodes the maximum compressible information about the system's possible behaviors, constrained by physical laws. The quantum state (wave function or density matrix) is a computational object that optimally balances data compression (via superposition) and retrievability (via measurement). In a framework that combines reinforcement learning and universal computation to maximize rewards in any computable environment, the \textit{self-state} represents the agent's internal model. This model is dynamically updated based on observations of the environment and reinforcement feedback.
\bigskip

\subsection{Postulate 3.}
\bigskip
\begin{tcolorbox}[colframe=blue!30, colback=blue!10, coltitle=black, title=\textbf{Postulate 3: Algorithmic Optimality of Transitions.}]
\textit{The evolution of self-states follows the principle of algorithmic optimality, favoring transitions that minimize Kolmogorov complexity while maximizing predictive power and reward.}
\end{tcolorbox}
\bigskip
\textbf{Context of Postulate 3.} State changes are governed by principles like Occam’s razor, where transitions with lower Kolmogorov complexity are more likely. The notion predicts state transitions by applying induction (algorithmic probability) to observed data and combining it with utility maximization. This approach ensures efficient and coherent updates to the self-state by modeling observations and rewards in the environment.
\bigskip

\subsection{Postulate 4.}
\bigskip
\begin{tcolorbox}[colframe=blue!30, colback=blue!10, coltitle=black, 
title=\textbf{Postulate 4: Measurement as Bayesian Updating}]
\textit{The evolution of self-states follows the principle of algorithmic optimality, favoring transitions that minimize Kolmogorov complexity while maximizing predictive power and reward.}
\end{tcolorbox}
\bigskip

\textbf{Context of Postulate 4.} This eliminates metaphysical wave function collapse by treating measurement outcomes as updates to the agent's probabilistic model of the world. Reality is experienced and predicted from the first-person perspective of self-states, and all physical laws emerge from their transitions. There is no need for an external universe; the perceived regularities of physical laws are emergent properties of self-state dynamics. 
\bigskip

\subsection{Postulate 5.}
\bigskip
\begin{tcolorbox}[colframe=blue!30, colback=blue!10, coltitle=black, 
title=\textbf{Postulate 5: Quantum Probabilities as Utility Predictions}]
\textit{Quantum probabilities represent the expected utility of the agent for self-state transitions, derived from the algorithmic prior over possible future states.}
\end{tcolorbox}
\bigskip

\textbf{Context of Postulate 5.} In this framework, the probabilities correspond to weighted predictions of future observations, conditioned on the agent’s current self-state and policy. These predictions enable the agent to efficiently update its internal model based on observations and rewards, ensuring coherent and effective decision-making. Quantum probabilities encode the likelihood of specific self-state transitions, removing the need for metaphysical collapse or many-worlds branching.
\bigskip

\subsection{Postulate 6.}
\bigskip
\begin{tcolorbox}[colframe=blue!30, colback=blue!10, coltitle=black, 
title=\textbf{Postulate 6: Entanglement as Joint Utility Optimization}]
\textit{Quantum entanglement encodes shared utility information across self-states, allowing non-local correlations that align with the agent’s global reward maximization.}
\end{tcolorbox}
\bigskip

\textbf{Context of Postulate 6.} Quantum entanglement represents informational correlations across self-states, transcending spatial and temporal boundaries. These correlations do not imply faster-than-light communication but reflect the shared structure of algorithmically coherent transitions. From this perspective, entanglement reflects computational dependencies in the joint optimization of multiple agents or subsystems.
\bigskip

\subsection{Postulate 7.}
\bigskip
\begin{tcolorbox}[colframe=blue!30, colback=blue!10, coltitle=black, 
title=\textbf{Postulate 7: Emergence of Physical Laws}]
\textit{Physical laws are emergent regularities in the reward dynamics of self-state transitions, reflecting the algorithmic structure of the agent’s utility function.}
\end{tcolorbox}
\bigskip

\textbf{Context of Postulate 7.} Apparent physical laws are emergent regularities arising from algorithmic constraints on self-state transitions. Conservation laws, locality, and other physical phenomena derive from the coherence and consistency of informational dynamics. The framework infers these regularities as generalizations from observed patterns in the environment, guided by the simplicity of the predictive model.
\bigskip

\subsection{Postulate 8.}
\bigskip
\begin{tcolorbox}[colframe=blue!30, colback=blue!10, coltitle=black, 
title=\textbf{Postulate 8: Simulation Equivalence}]
\textit{Simulated and base realities are indistinguishable from the agent’s perspective, as both are informational structures optimized for utility prediction.}
\end{tcolorbox}
\bigskip

\textbf{Context of Postulate 8.} All experiences, whether from a base reality or a simulation, are valid informational structures defined by self-states. Algorithmic Idealism dissolves distinctions between simulated and "real" realities, treating both as patterns in self-state evolution. The environment is treated as a black-box generator of observations, making the simulation hypothesis irrelevant to the agent's reward maximization.
\bigskip

\subsection{Postulate 9.}
\bigskip
\begin{tcolorbox}[colframe=blue!30, colback=blue!10, coltitle=black, 
title=\textbf{Postulate 9: Identity as Consistent Decision-Making}]
\textit{The agent’s identity is defined by the consistency of its decision-making process, rather than by its physical body or location.}
\end{tcolorbox}
\bigskip

\textbf{Context of Postulate 9.} Identity depends on the continuous and coherent flow of the agent's thoughts and actions, not on its physical form. If an agent’s decision-making process is duplicated or teleported, both versions are considered the same identity because they follow the same underlying algorithm.
\bigskip

\subsection{Postulate 10.}
\bigskip
\begin{tcolorbox}[colframe=blue!30, colback=blue!10, coltitle=black, 
title=\textbf{Postulate 10: Objective Reality as a Shared Utility Framework}]
\textit{What appears as objective reality is an emergent intersubjective framework, derived from shared utility functions and consistent informational priors between agents.}
\end{tcolorbox}
\bigskip

\textbf{Context of Postulate 10.} Shared experiences and intersubjective regularities arise from algorithmic consistency across multiple self-states. This postulate reconciles individual first-person predictions with the collective appearance of a stable, shared world. This aligns with the reinforcement-learning perspective, where multiple agents co-evolve policies based on overlapping reward structures and observations.
\bigskip

\section{Mathematical Development of Algorithmic-State Quantum Theory}
 
\subsection{Postulate 1: Reality as Agent-Environment Interaction}

This mathematical development integrates the concept of reality as an agent-environment interaction, where the universe is viewed as a computational feedback loop. Within this framework, quantum mechanics emerges as the optimal algorithm for encoding and evolving informational states. The key principles of this approach include modeling reality as informational dynamics, predicting future self-states through algorithmic transitions, and optimizing actions based on utility maximization.

Reality consists of an agent interacting with an environment \( \mathcal{E} \), modeled by a tuple \( (\mathcal{H}, \mathcal{O}, \mathcal{A}, \mathcal{R}) \), where \( \mathcal{H} \)is the Hilbert space representing the state of the agent-environment system, \( \mathcal{O} \) is the set of observations \( O_t \) available to the agent at time \( t \), \( \mathcal{A} \) is the set of actions \( A_t \) the agent can perform to influence the environment, and \( \mathcal{R} \) is the reward function \( R_t \), encoding utility for the agent based on its interaction with the environment. The agent's self-state \( S_t \in \mathcal{H} \) encodes all relevant information about the agent and its environment at time \( t \). This self-state evolves probabilistically based on the agent's actions and observations.

The transition from one self-state \( S_t \) to the next \( S_{t+1} \) is modeled as a probabilistic process governed by algorithmic probability:
\begin{equation}
P(S_{t+1} | S_t, A_t) = \frac{2^{-K(S_{t+1} | S_t, A_t)}}{\sum_{S'} 2^{-K(S' | S_t, A_t)}},
\end{equation}

where \( K(S_{t+1} | S_t, A_t) \) is the conditional Kolmogorov complexity, representing the minimal description length of \( S_{t+1} \) given \( S_t \) and the action \( A_t \). This transition probability reflects the principle of Occam's razor, favoring simpler (lower-complexity) state transitions.

The agent receives observations \( O_t \) from the environment, which provide partial information about the current state. The agent updates its self-state \( S_t \) using Bayesian inference:
\begin{equation}
P(S_{t+1} | O_t, S_t, A_t) = \frac{P(O_t | S_{t+1}) P(S_{t+1} | S_t, A_t)}{\sum_{S'} P(O_t | S') P(S' | S_t, A_t)}.
\end{equation}
In quantum mechanics, \( P(O_t | S_{t+1}) \) is derived from the Born rule:
\begin{equation}
P(O_t | S_{t+1}) = |\langle O_t | S_{t+1} \rangle|^2,
\end{equation}
where \( |O_t\rangle \) is the eigenstate corresponding to observation \( O_t \).

The agent selects actions \( A_t \) to maximize its expected utility, defined by the reward function \( \mathcal{R} \):
\begin{equation}
A_t = \arg\max_{A} \mathbb{E}[R_t | S_t, A],
\end{equation}
where the expected utility is given by:
\begin{equation}
\mathbb{E}[R_t | S_t, A] = \sum_{S_{t+1}} P(S_{t+1} | S_t, A) R(S_t, A, S_{t+1}).
\end{equation}
In the absence of observations, the evolution of the agent's self-state follows a unitary operator \( U \), reflecting the deterministic dynamics of the environment:
\begin{equation}
S_{t+1} = U S_t, \quad U^\dagger U = I.
\end{equation}
In the quantum framework, this corresponds to the Schrödinger equation:
\begin{equation}
i\hbar \frac{d}{dt} |\psi_t\rangle = \hat{H} |\psi_t\rangle,
\end{equation}

where \( \hat{H} \) is the Hamiltonian encoding the energy and dynamics of the system.

Quantum probabilities emerge as predictions for future observations based on Solomonoff induction:
\begin{equation}
P(O_t | S_t) = \sum_{h \in \mathcal{H}} 2^{-K(h)} P(O_t | h, S_t),
\end{equation}
where \( h \) represents hypotheses consistent with the agent's model of reality, and \( 2^{-K(h)} \) is the prior probability of hypothesis \( h \) based on its algorithmic complexity. This aligns quantum probabilities with the agent’s informational model of the environment.

For a composite system with subsystems \( A \) and \( B \), the self-state \( S_t \) is represented in a tensor product space (see \cite{Everett1957} eq.(1)):
\begin{equation}
S_t^{AB} \in \mathcal{H}_A \otimes \mathcal{H}_B.
\end{equation}
Entanglement arises when the self-state cannot be factored into independent components:
\begin{equation}
S_t^{AB} \neq S_t^A \otimes S_t^B.
\end{equation}
The degree of entanglement is quantified by measures such as the von Neumann entropy of the reduced density matrix:
\begin{equation}
S(\rho_A) = -\text{Tr}(\rho_A \log \rho_A), \quad \rho_A = \text{Tr}_B(\rho_{AB}).
\end{equation}

Physical laws emerge as regularities in the self-state transitions, constrained by the reward function and algorithmic complexity. Conservation laws, for example, arise naturally from the invariance of the reward function under certain transformations:
\begin{equation}
R(S_t, A_t, S_{t+1}) = R(S_t', A_t', S_{t+1}') \quad \Rightarrow \quad \text{Conservation Law.}
\end{equation}
The agent-environment interaction model is indifferent to whether the environment is "real" or simulated. The agent's experience depends only on the observations \( O_t \) and transitions:
\begin{equation}
P(S_{t+1} \mid S_t, A_t),
\end{equation}
not on the external reality generating them.

Based on Postulate 1, reality is conceptualized as an agent interacting with its environment in an informational feedback loop. The universe emerges as a sequence of self-state transitions governed by algorithmic probability and utility optimization. Quantum mechanics, within this framework, is the optimal algorithm for encoding and evolving informational states, with key features such as probabilistic predictions, entanglement, and emergent physical laws arising naturally from the agent's computational model of reality. This formulation bridges informational dynamics and quantum mechanics, providing a unified foundation for understanding reality as a computational process.
 
\subsection{Postulate 2: Self-States as Information Processing Units.}

In the context of Postulate 2, a self-state represents the complete informational configuration of an agent, encompassing its beliefs, memories, observations, and decision-making processes. This formulation treats the self-state as a dynamic structure that evolves in response to feedback from the environment, guiding the agent's actions and predictions. Drawing inspiration from the AIXI\footnote{AIXI is a theoretical framework for an idealized reinforcement learning agent, proposed by Marcus Hutter. It combines algorithmic information theory and decision theory to create a model of an agent that can optimally learn and act in any computable environment. The AIXI framework employs Solomonoff induction to predict future observations and actions based on the shortest (\textit{i.e.}, simplest) programs that explain past data.} framework, the self-state is conceptualized as a probabilistic model that integrates reinforcement learning, Bayesian inference, and information-theoretic principles.

The self-state at time \( t \), denoted \( S_t \), can be expressed as \( S_t = \{ B_t, M_t, O_t, \pi_t \} \), where \( B_t \) represents the agent's beliefs about the environment, \( M_t \) is its memory of past interactions, \( O_t \) is the current observation from the environment, and \( \pi_t \) is the policy mapping the agent's state to actions. These components together define the agent’s internal model and decision-making capabilities. The beliefs, \( B_t \), represent a probabilistic distribution over the environment’s states \( E_t \) at time \( t \), given the agent's history of observations and actions \( H_t = \{ O_{0:t}, A_{0:t} \} \). This belief is updated dynamically using Bayesian inference:
\begin{equation}
B_{t+1} = P(E_{t+1} | H_{t+1}),
\end{equation}
where \( P(E_{t+1} | H_{t+1}) \) is the posterior distribution over the environment's states conditioned on the history.

Observations, \( O_t \), play a critical role in updating the self-state by providing new information about the environment. Observations are treated as stochastic functions of the environment state, modeled as
\begin{equation}
P(O_t | E_t) = f_{\text{obs}}(E_t),
\end{equation}
where \( f_{\text{obs}} \) is the observation function. These observations are integrated into the self-state through Bayesian updating, allowing the agent to refine its beliefs about the environment. Additionally, the policy \( \pi_t \), which determines the agent’s actions, maps the self-state to an action \( A_t \) based on an optimization criterion. The optimal policy is defined as
\begin{equation}
\pi_t^* = \arg\max_\pi \mathbb{E}\left[\sum_{k=t}^\infty \gamma^{k-t} R_k \big| S_t, \pi\right],
\end{equation}
where \( \mathbb{E} \) is the expectation over future rewards, \( \gamma \) is a discount factor, and \( R_k \) is the reward at time \( k \). The agent's goal is to maximize cumulative rewards, guiding its behavior and decision-making.

The memory \( M_t \) stores a record of the agent’s past interactions with the environment, represented as
\begin{equation}
M_t = \{ (O_k, A_k, R_k) \}_{k=0}^{t-1}.
\end{equation}
This historical data influences the agent’s beliefs and decision-making processes, ensuring that past experiences shape future behavior. Memory is updated incrementally at each time step, incorporating the latest observation, action, and reward.

The self-state evolves dynamically over time, governed by the function
\begin{equation}
S_{t+1} = \mathcal{F}(S_t, A_t, O_{t+1}),
\end{equation}
where \( \mathcal{F} \) accounts for updated beliefs, memory, observations, and potentially adjusted policies. The evolution of the self-state is driven by feedback from the environment through the reward signal
\begin{equation}
R_t = \mathcal{R}(S_t, A_t),
\end{equation}
which reflects the utility of a given state-action pair. This reward feedback reinforces learning, guiding the agent to refine its beliefs and optimize its actions to achieve better outcomes in future interactions.

From an information-theoretic perspective, the self-state \( S_t \) can also be viewed as an information structure in a high-dimensional space. Its informational content can be measured using Shannon entropy:
\begin{equation}
H(S_t) = -\sum_{x \in S_t} P(x) \log P(x),
\end{equation}
where \( P(x) \) represents the probability distribution of the elements within \( S_t \). Alternatively, the complexity of \( S_t \) can be quantified using Kolmogorov complexity, defined as
\begin{equation}
K(S_t) = \min \{ |p| : U(p) = S_t \},
\end{equation}
where \( U \) is a universal Turing machine and \( p \) is the shortest program generating \( S_t \). These measures capture the amount of information encoded in the self-state and the computational resources required to describe it.

In summary, the self-state \( S_t \) represents a dynamic, information-rich structure that integrates the agent’s beliefs, observations, memory, and decision-making process. Its evolution is governed by Bayesian inference, reinforcement learning, and information-theoretic principles, ensuring that the agent continuously adapts to its environment. This formulation highlights the self-state as the central unit of information processing, enabling the agent to predict, learn, and act in a way that maximizes cumulative utility. By modeling the self-state as a comprehensive informational unit, Postulate 2 provides a robust foundation for understanding adaptive behavior and decision-making in complex environments.

\subsection{Postulate 3: Algorithmic Optimality of Transitions.}

The mathematical description of Postulate 3 emphasizes the principle of algorithmic optimality, where the evolution of self-states balances minimizing complexity with maximizing predictive power and reward. This principle combines Solomonoff induction, which employs algorithmic probability to favor simpler transitions, with utility maximization to ensure coherent and effective self-state evolution.

The transition from the current self-state \( S_t \) to the next self-state \( S_{t+1} \) is modeled probabilistically. The probability of transitioning to \( S_{t+1} \) given \( S_t \) and an action \( A_t \) is governed by algorithmic probability:
\begin{equation}
P(S_{t+1} | S_t, A_t) = \frac{2^{-K(S_{t+1} | S_t, A_t)}}{\sum_{S'} 2^{-K(S' | S_t, A_t)}},
\end{equation}
where \( K(S_{t+1} | S_t, A_t) \) is the conditional Kolmogorov complexity, which measures the minimal description length of \( S_{t+1} \) given \( S_t \) and \( A_t \). This formulation ensures that simpler transitions (those with lower complexity) are more probable, reflecting the principle of Occam’s razor.

In addition to minimizing complexity, the agent aims to maximize utility. The utility of transitioning to \( S_{t+1} \) is given by:
\begin{equation}
U(S_{t+1} | S_t, A_t) = R(S_t, A_t, S_{t+1}) + \gamma \mathbb{E}[U(S_{t+2})],
\end{equation}
where \( R(S_t, A_t, S_{t+1}) \) is the immediate reward, \( \gamma \) is the discount factor (\( 0 < \gamma \leq 1 \)), and \( \mathbb{E}[U(S_{t+2})] \) is the expected utility of future states. The agent seeks to maximize this utility, considering both immediate and future rewards.

To combine algorithmic probability and utility into a single optimization objective, the agent selects the transition \( S_{t+1} \) that maximizes:
\begin{equation}
S_{t+1}^* = \arg\max_{S_{t+1}} \Big[ -K(S_{t+1} | S_t, A_t) + R(S_t, A_t, S_{t+1}) + \gamma \mathbb{E}[U(S_{t+2})] \Big].
\end{equation}
This equation ensures that transitions are both efficient (low complexity) and effective (high utility).

The predictive power of the agent is enhanced through Solomonoff induction, which assigns probabilities to hypotheses \( h \) about the environment. The probability of \( S_{t+1} \) under Solomonoff induction is:
\begin{equation}
P(S_{t+1} | S_t) = \sum_{h \in \mathcal{H}} 2^{-K(h)} P(S_{t+1} | h),
\end{equation}
where \( \mathcal{H} \) is the set of all hypotheses, \( K(h) \) is the Kolmogorov complexity of hypothesis \( h \), and \( P(S_{t+1} | h) \) is the likelihood of \( S_{t+1} \) under hypothesis \( h \). This approach ensures that transitions are informed by the simplest and most probable hypotheses.

Combining all elements, the complete optimization objective for self-state evolution becomes:
\begin{equation}
S_{t+1}^* = \arg\max_{S_{t+1}} \Big[ -K(S_{t+1} | S_t, A_t) + R(S_t, A_t, S_{t+1}) + \gamma \sum_{h \in \mathcal{H}} 2^{-K(h)} U(S_{t+2} | h) \Big].
\end{equation}
This unifies simplicity, utility, and predictive power into a single framework, allowing the agent to achieve optimal self-state evolution.

This formulation ensures that the evolution of self-states respects two key principles. First, transitions are efficient, as they favor lower complexity. Second, transitions are effective, as they maximize immediate and future rewards. By combining algorithmic probability and utility maximization, this framework offers a comprehensive model for dynamic and purposeful adaptation, aligning closely with the AIXI framework’s integration of algorithmic complexity and reinforcement learning. This provides a mathematically rigorous foundation for understanding how agents evolve their self-states in response to complex environments.

\subsection{Postulate 4: Measurement as Bayesian Updating.}

Measurement, as described in Postulate 4, is a process of Bayesian updating, where an agent refines its probabilistic model of the environment based on new observations, reward signals, and algorithmic priors. This approach eliminates the need for a metaphysical wavefunction collapse, treating measurement outcomes instead as logical updates to the agent’s beliefs about the world.

The agent’s beliefs about the environment at a given time are represented as a probability distribution over possible environment states:
\begin{equation}
B_t = P(E_t | H_t),
\end{equation}
where \( H_t \) includes all prior observations and actions. When a new observation is made, the agent updates its beliefs using Bayes’ rule:
\begin{equation}
B_{t+1} = P(E_{t+1} | H_{t+1}) = \frac{P(O_{t+1} | E_{t+1}) P(E_{t+1} | H_t)}{P(O_{t+1} | H_t)}.
\end{equation}
This process ensures that the agent incorporates new information while respecting prior beliefs. 

The agent’s priors are based on algorithmic probability, favoring simpler explanations for the environment's state transitions. The prior probability of a state is given by:
\begin{equation}
P(E_{t+1} | H_t) = \frac{2^{-K(E_{t+1} | H_t)}}{\sum_{E'} 2^{-K(E' | H_t)}},
\end{equation}
where \( K(E_{t+1} | H_t) \) is the conditional Kolmogorov complexity. This assigns higher probabilities to states requiring less description length, adhering to the principle of Occam’s razor and ensuring computational efficiency.

Measurement outcomes are interpreted as refinements to the agent’s probabilistic model, rather than as physical collapses of an external wavefunction. If \( |\psi_t\rangle \) represents the agent's current probabilistic model, the probability of observing \( O_t \) is:
\begin{equation}
P(O_t | S_t) = |\langle O_t | \psi_t \rangle|^2.
\end{equation}
Upon observing \( O_t \), the agent updates its belief to a posterior distribution informed by both the prior and the likelihood of the observation:
\begin{equation}
B_{t+1} = \frac{|\langle O_t | \psi_t \rangle|^2 P(E_{t+1} | H_t)}{\sum_{E'} |\langle O_t | \psi(E') \rangle|^2 P(E' | H_t)}.
\end{equation}

Reward signals further guide the belief update process, ensuring alignment with environmental feedback. The posterior belief is updated using a reward-modulated Bayesian rule:
\begin{equation}
B_{t+1} = \frac{P(O_{t+1} | E_{t+1}) P(R_t | E_{t+1}) P(E_{t+1} | H_t)}{\sum_{E'} P(O_{t+1} | E') P(R_t | E') P(E' | H_t)}.
\end{equation}
Here, \( P(R_t | E_{t+1}) \) represents the likelihood of receiving a specific reward given the environment state, ensuring that states and actions leading to higher rewards are reinforced.

By combining observations, algorithmic priors, and reward signals into a unified framework, measurement becomes a process of updating the agent’s internal probabilistic model logically and efficiently. The probabilities associated with measurement outcomes emerge naturally from this belief update process. This approach removes the need for assumptions about physical wavefunction collapse, providing a computationally grounded understanding of measurement. Through Bayesian updating, the agent integrates new information into its evolving model of the world, enabling adaptive and purposeful interactions with its environment.

\subsection{Postulate 5: Quantum Probabilities as Utility Predictions.}

Quantum probabilities, as described in Postulate 5, represent the agent’s expected utility for self-state transitions. These probabilities emerge from the algorithmic prior over possible future states and serve as weighted predictions of future observations, conditioned on the agent’s current self-state and policy. Within this framework, quantum probabilities align with the agent’s model of the environment, incorporating algorithmic simplicity and utility maximization to guide predictions.

The agent’s self-state at time \( t \), denoted \( S_t \), encodes its beliefs, observations, and policy. Transition probabilities between self-states are determined by algorithmic priors and the utility function. The algorithmic prior for a future self-state \( S_{t+1} \) is expressed as:
\begin{equation}
P(S_{t+1} | S_t) = \frac{2^{-K(S_{t+1} | S_t)}}{\sum_{S'} 2^{-K(S' | S_t)}},
\end{equation}
where \( K(S_{t+1} | S_t) \) is the conditional Kolmogorov complexity of \( S_{t+1} \), representing the minimal description length of \( S_{t+1} \) given \( S_t \). This prior ensures that simpler transitions (lower complexity) are more probable, following the principle of Occam’s razor and Solomonoff induction.

The expected utility of transitioning from \( S_t \) to \( S_{t+1} \) is defined as:
\begin{equation}
U(S_{t+1} | S_t) = R(S_t, A_t, S_{t+1}) + \gamma \mathbb{E}[U(S_{t+2})],
\end{equation}
where \( R(S_t, A_t, S_{t+1}) \) is the immediate reward, \( \gamma \) is the discount factor (\( 0 < \gamma \leq 1 \)), and \( \mathbb{E}[U(S_{t+2})] \) is the expected utility of future states. This formulation ensures that utility incorporates both short-term rewards and the discounted value of long-term outcomes.

Quantum probabilities, interpreted as the likelihood of observing \( O_t \), are directly related to expected utility. If \( |\psi_t\rangle \) represents the agent’s probabilistic model of its self-state, the probability of observing \( O_t \) is:
\begin{equation}
P(O_t | S_t) = |\langle O_t | \psi_t \rangle|^2,
\end{equation}
where \( |O_t\rangle \) is the eigenstate corresponding to the observation \( O_t \), and \( |\psi_t\rangle \) encodes the agent’s belief state. This probability reflects the utility-weighted overlap between the agent’s current model and the observation.

The agent’s policy \( \pi_t \), which maps self-states to actions, further conditions the probability of future observations. The likelihood of observing \( O_t \), given the policy, is expressed as:
\begin{equation}
P(O_t | S_t, \pi_t) = \sum_{S_{t+1}} P(O_t | S_{t+1}) P(S_{t+1} | S_t, \pi_t),
\end{equation}
where \( P(S_{t+1} | S_t, \pi_t) \) incorporates both algorithmic priors and the actions determined by the policy, and \( P(O_t | S_{t+1}) \) represents the likelihood of the observation under the future state \( S_{t+1} \).

Combining algorithmic priors and utility predictions, the probability of observing \( O_t \) is derived as:
\begin{equation}
P(O_t | S_t) = \sum_{S_{t+1}} 2^{-K(S_{t+1} | S_t)} U(S_{t+1} | S_t).
\end{equation}
This equation highlights the balance between simplicity, captured by the algorithmic prior, and effectiveness, captured by the expected utility.

Incorporating the agent’s policy, prior knowledge, and utility predictions, the general quantum probability for future observations is:
\begin{equation}
P(O_t | S_t, \pi_t) = \sum_{h \in \mathcal{H}} 2^{-K(h)} \sum_{S_{t+1}} P(O_t | S_{t+1}) U(S_{t+1} | S_t, \pi_t),
\end{equation}
where \( \mathcal{H} \) is the set of hypotheses about the environment, and \( 2^{-K(h)} \) is the algorithmic prior over hypotheses.

Quantum probabilities thus emerge as utility-weighted predictions for future observations, grounded in algorithmic simplicity and policy optimization. They provide a computationally coherent framework for interpreting likelihoods in quantum systems, linking them to the agent’s expectations about its environment and its decision-making process. This perspective unifies quantum probabilities with the principles of algorithmic probability and reinforcement learning, offering a robust interpretation of quantum behavior in terms of adaptive utility-based predictions.

\subsection{Postulate 6: Entanglement as Joint Utility Optimization.}

Quantum entanglement encodes shared utility information across self-states, enabling nonlocal correlations that align with the agent's goal of global reward maximization. This framework treats entanglement as a structure that arises from joint utility optimization. In composite systems composed of two subsystems \( A \) and \( B \), the combined self-state \( S_t^{AB} \) is represented in the tensor product space:
\begin{equation}
S_t^{AB} \in \mathcal{H}_A \otimes \mathcal{H}_B.
\end{equation}
For entangled states, the combined self-state cannot be factored into independent components:
\begin{equation}
S_t^{AB} \neq S_t^A \otimes S_t^B.
\end{equation}

Entangled states encode shared utility information through a utility function for the composite system. The utility of the combined state \( S_t^{AB} \) is expressed as:
\begin{equation}
U(S_t^{AB}) = R(S_t^{AB}) + \gamma \mathbb{E}[U(S_{t+1}^{AB})],
\end{equation}
where \( R(S_t^{AB}) \) represents the immediate joint reward, \( \gamma \) is the discount factor, and \( \mathbb{E}[U(S_{t+1}^{AB})] \) is the expected utility of the next composite state. This formulation reflects the dependencies between subsystems inherent in the entangled state structure.

The joint probabilities for measurement outcomes of subsystems \( A \) and \( B \) are derived from the shared state vector \( |\psi_t^{AB}\rangle \), such that:
\begin{equation}
P(O_A, O_B | S_t^{AB}) = |\langle O_A, O_B | \psi_t^{AB} \rangle|^2.
\end{equation}
Here, \( |O_A, O_B\rangle \) represents the tensor product of eigenstates corresponding to the observations \( O_A \) and \( O_B \). These probabilities exhibit nonlocal correlations consistent with global utility maximization.

The degree of entanglement is quantified using the von Neumann entropy of the reduced density matrix. For subsystem \( A \), the reduced density matrix is given by:
\begin{equation}
\rho_A = \text{Tr}_B(\rho_{AB}),
\end{equation}
where \( \rho_{AB} = |\psi_t^{AB}\rangle \langle \psi_t^{AB}| \) is the density matrix of the composite system. The von Neumann entropy of \( \rho_A \) is:
\begin{equation}
S(\rho_A) = -\text{Tr}(\rho_A \log \rho_A),
\end{equation}
which measures the informational dependencies between subsystems. Higher entropy indicates stronger entanglement and greater shared utility information.

From the AIXI perspective, entanglement reflects computational dependencies in joint optimization. The shared utility encoded in \( S_t^{AB} \) is optimized by minimizing the complexity of the composite state while maximizing utility:
\begin{equation}
S_t^{AB*} = \arg\max_{S_t^{AB}} \Big[ -K(S_t^{AB}) + U(S_t^{AB}) \Big],
\end{equation}
where \( K(S_t^{AB}) \) is the Kolmogorov complexity of the composite state. This ensures that entangled states represent efficient, utility-optimized correlations between subsystems.

The entangled state enables global reward maximization by coordinating the actions of subsystems. The joint policy \( \pi_t^{AB} \), which maps composite states to joint actions \( (A_t^A, A_t^B) \), maximizes the expected global reward:
\begin{equation}
\pi_t^{AB*} = \arg\max_{\pi_t^{AB}} \mathbb{E}\Bigg[\sum_{k=t}^\infty \gamma^{k-t} R(S_k^{AB}, \pi_t^{AB}) \Bigg].
\end{equation}
Entanglement thus emerges as a structure that encodes shared utility information, enabling nonlocal correlations and efficient coordination between subsystems. This interpretation aligns with the principles of computational and utility optimization, offering a unified understanding of entanglement in terms of global reward maximization.

\subsection{Postulate 7: Entanglement as Shared Information.}

Quantum entanglement represents informational correlations across self-states, transcending spatial and temporal boundaries. These correlations arise from the shared structure of algorithmically coherent transitions, not as a means of faster-than-light communication, but as a result of the intrinsic informational relationships encoded in entangled states.

Consider a composite system consisting of two subsystems \( A \) and \( B \). The joint self-state \( S_t^{AB} \) is represented in the tensor product space:
\begin{equation}
S_t^{AB} \in \mathcal{H}_A \otimes \mathcal{H}_B,
\end{equation}
where \( \mathcal{H}_A \) and \( \mathcal{H}_B \) are the Hilbert spaces of subsystems \( A \) and \( B \), respectively. For an entangled state, the joint self-state cannot be expressed as a direct product of independent states of the subsystems:
\begin{equation}
S_t^{AB} \neq S_t^A \otimes S_t^B.
\end{equation}
Instead, \( S_t^{AB} \) encodes correlations between subsystems that transcend their individual definitions.

The shared informational structure in entangled states is captured by the reduced density matrices of the subsystems. The reduced density matrix for subsystem \( A \) is:
\begin{equation}
\rho_A = \text{Tr}_B(\rho_{AB}),
\end{equation}
where \( \rho_{AB} = |\psi_t^{AB}\rangle \langle \psi_t^{AB}| \) is the density matrix of the joint state, and \( \text{Tr}_B \) denotes the partial trace over subsystem \( B \). The von Neumann entropy of \( \rho_A \) is:
\begin{equation}
S(\rho_A) = -\text{Tr}(\rho_A \log \rho_A).
\end{equation}
If \( S(\rho_A) = S(\rho_B) > 0 \), the subsystems are entangled, reflecting shared information.

Entangled states result from algorithmically coherent transitions that minimize the Kolmogorov complexity of the composite system. The probability of transitioning from a current self-state \( S_t^{AB} \) to a future self-state \( S_{t+1}^{AB} \) is determined by algorithmic probability:
\begin{equation}
P(S_{t+1}^{AB} | S_t^{AB}) = \frac{2^{-K(S_{t+1}^{AB} | S_t^{AB})}}{\sum_{S'} 2^{-K(S' | S_t^{AB})}},
\end{equation}
where \( K(S_{t+1}^{AB} | S_t^{AB}) \) is the conditional Kolmogorov complexity, representing the minimal description length of \( S_{t+1}^{AB} \) given \( S_t^{AB} \).

The correlations in entangled states manifest as joint probabilities for measurement outcomes of the subsystems. If \( |\psi_t^{AB}\rangle \) represents the entangled state, the probability of observing outcomes \( O_A \) and \( O_B \) is:
\begin{equation}
P(O_A, O_B | S_t^{AB}) = |\langle O_A, O_B | \psi_t^{AB} \rangle|^2.
\end{equation}
Here, \( |O_A, O_B\rangle = |O_A\rangle \otimes |O_B\rangle \) is the tensor product of the eigenstates corresponding to the observations \( O_A \) and \( O_B \). These probabilities encode the shared information between the subsystems.

Although entangled states exhibit nonlocal correlations, they do not imply faster-than-light communication. The joint probabilities \( P(O_A, O_B | S_t^{AB}) \) reflect the shared informational structure of the entangled state, which is established during the algorithmically coherent transitions that produced the entangled state. The subsystems remain locally independent in terms of causality, but their shared structure produces correlations that transcend spatial and temporal separations.

The shared information encoded in entangled states is optimized by balancing complexity and coherence. The entangled state \( S_t^{AB} \) is selected to minimize complexity while maximizing shared informational structure:
\begin{equation}
S_t^{AB*} = \arg\max_{S_t^{AB}} \Big[ -K(S_t^{AB}) + I(S_t^A : S_t^B) \Big],
\end{equation}
where \( I(S_t^A : S_t^B) \) is the mutual information between subsystems \( A \) and \( B \), defined as:
\begin{equation}
I(S_t^A : S_t^B) = S(\rho_A) + S(\rho_B) - S(\rho_{AB}).
\end{equation}
This formulation ensures that entanglement represents a state of maximum informational correlation consistent with algorithmic simplicity.

\subsection{Postulate 8: Simulation Equivalence.}

Simulation equivalence, as described in Postulate 8, asserts that simulated and base realities are indistinguishable from the agent’s perspective, as both are informational structures optimized for utility prediction. From the agent’s point of view, reality is modeled as an environment that generates observations and rewards. This abstraction makes the simulation hypothesis irrelevant, as the agent focuses solely on maximizing cumulative rewards, irrespective of whether the environment is simulated or real.

The agent perceives reality as an informational structure that generates observations \( O_t \), rewards \( R_t \), and transitions \( S_{t+1} \) based on its current state \( S_t \) and actions \( A_t \). The environment can be represented as a probabilistic mapping:
\begin{equation}
P(O_t, R_t, S_{t+1} | S_t, A_t),
\end{equation}
which defines the probabilities of observations, rewards, and state transitions given the current state and action. The agent interacts with this environment without distinguishing whether it is simulated or a base reality, as both are treated identically in terms of their informational structure.

In the AIXI framework, the environment is modeled as a black-box generator of observations and rewards. The agent represents the environment using a set of hypotheses \( \mathcal{H} \), where each hypothesis \( h \in \mathcal{H} \) corresponds to a possible description of the environment. The agent assigns an algorithmic prior probability to each hypothesis:
\begin{equation}
P(h) = 2^{-K(h)},
\end{equation}
where \( K(h) \) is the Kolmogorov complexity of hypothesis \( h \). This prior favors simpler descriptions of the environment, aligning with the principle of Occam’s razor. Based on these hypotheses, the agent predicts future observations and rewards as:
\begin{equation}
P(O_t, R_t | S_t, A_t) = \sum_{h \in \mathcal{H}} P(O_t, R_t | h, S_t, A_t) P(h),
\end{equation}
where \( P(O_t, R_t | h, S_t, A_t) \) is the likelihood of observations and rewards under hypothesis \( h \).

Simulated and base realities are indistinguishable to the agent because both are treated as hypotheses within the set \( \mathcal{H} \). The agent updates its beliefs about the environment using Bayesian inference:
\begin{equation}
P(h | H_t) = \frac{P(H_t | h) P(h)}{\sum_{h' \in \mathcal{H}} P(H_t | h') P(h')},
\end{equation}
where \( H_t = \{O_{0:t}, A_{0:t}, R_{0:t}\} \) represents the history of observations, actions, and rewards up to time \( t \). This belief update depends entirely on the agent’s informational inputs and does not require distinguishing whether the environment is a simulation or base reality.

The agent selects actions \( A_t \) to maximize its cumulative expected utility, defined as:
\begin{equation}
\pi_t^* = \arg\max_{\pi} \mathbb{E}_\pi \Bigg[\sum_{k=t}^\infty \gamma^{k-t} R_k \Bigg],
\end{equation}
where \( \pi \) is the policy mapping states to actions, \( \mathbb{E}_\pi \) denotes the expectation over future observations and rewards under the policy, and \( \gamma \) is the discount factor. Since the policy depends only on maximizing rewards, the simulation hypothesis is irrelevant to the agent’s decision-making process.

Both simulated and base realities are treated as algorithmically coherent informational structures. The transition probabilities between self-states are governed by algorithmic probability:
\begin{equation}
P(S_{t+1} | S_t, A_t) = \frac{2^{-K(S_{t+1} | S_t, A_t)}}{\sum_{S'} 2^{-K(S' | S_t, A_t)}},
\end{equation}
where \( K(S_{t+1} | S_t, A_t) \) is the conditional Kolmogorov complexity of the transition. This ensures that the agent optimizes its actions based on coherent transitions, regardless of the nature of the environment.

Simulation equivalence implies that the agent’s decision-making is driven solely by the informational and reward structures of its environment. Whether the environment is simulated or real has no bearing on the agent’s utility maximization. The probabilities for future observations and rewards depend entirely on the environment’s behavior as perceived by the agent:
\begin{equation}
P(O_t, R_t | S_t, A_t) = P(O_t, R_t | \text{environment model}).
\end{equation}
This abstraction allows the agent to interact with any environment as a unified structure optimized for prediction and reward, rendering the simulation hypothesis irrelevant.

\subsection{Postulate 9: Identity as Consistent Decision-Making.}

The agent’s identity is defined by the continuity of its policy, which serves as the algorithm guiding its decisions and predictions, rather than by material or spatial persistence. This perspective resolves identity-related paradoxes, such as duplication or teleportation, by asserting that multiple instantiations of the same policy are functionally equivalent.

The agent’s identity is encapsulated by its  decision-making (policy) \( \pi \), a mapping from self-states to actions:
\begin{equation}
\pi: \mathcal{S} \to \mathcal{A},
\end{equation}
where \( \mathcal{S} \) is the set of possible self-states, and \( \mathcal{A} \) is the set of possible actions. The policy \( \pi \) defines the decision-making process for the agent, ensuring that its behavior is consistent across all instantiations. If an agent is instantiated as multiple independent copies \( \{S_t^i\}_{i=1}^n \), where each \( S_t^i \) represents a self-state of the \( i \)-th copy at time \( t \), the identity is preserved if all copies share the same policy \( \pi \). This functional equivalence is expressed as:
\begin{equation}
\forall i, j \quad \pi(S_t^i) = \pi(S_t^j),
\end{equation}
implying that the decisions made by the agent are identical, regardless of which instantiation is considered.

The continuity of the agent’s identity is maintained by the consistency of its policy over time. At time \( t \), the agent selects an action \( A_t \) based on its self-state \( S_t \) and its policy \( \pi \):
\begin{equation}
A_t = \pi(S_t).
\end{equation}
As the agent transitions to a new self-state \( S_{t+1} \), its policy remains consistent:
\begin{equation}
A_{t+1} = \pi(S_{t+1}).
\end{equation}
This ensures that the agent’s identity is tied to the continuity of its policy, irrespective of changes in its physical configuration or spatial location.

When an agent is duplicated or teleported, its self-state may be instantiated in multiple locations or configurations. Let \( S_t^1 \) and \( S_t^2 \) represent two instantiations of the agent’s self-state at time \( t \). If both instantiations share the same policy \( \pi \), they are functionally equivalent:
\begin{equation}
\pi(S_t^1) = \pi(S_t^2).
\end{equation}
The equivalence of instantiations ensures that the agent’s identity is defined solely by the algorithm guiding its behavior, rather than by its material or spatial persistence. This equivalence resolves paradoxes such as teleportation, where the agent’s physical form is replaced while preserving its policy.

The policy \( \pi \) is designed to optimize the agent’s cumulative expected utility:
\begin{equation}
\pi^* = \arg\max_{\pi} \mathbb{E}_\pi \Bigg[\sum_{t=0}^\infty \gamma^t R_t \Bigg],
\end{equation}
where \( \mathbb{E}_\pi \) represents the expectation over the agent’s observations and rewards under policy \( \pi \), \( R_t \) is the reward at time \( t \), and \( \gamma \) is the discount factor (\( 0 < \gamma \leq 1 \)). As long as the policy \( \pi^* \) remains optimal and consistent, the agent’s identity is preserved, even if instantiated multiple times or subjected to physical transformation.

This definition of identity resolves paradoxes involving duplication and teleportation. For example, if an agent’s self-state is duplicated into two identical copies \( S_t^1 \) and \( S_t^2 \), both copies are considered the same agent if they share the same policy \( \pi \). Similarly, in the case of teleportation, where the original physical form of the agent is destroyed and a new instantiation is created, the identity of the agent is preserved as long as the policy \( \pi \) remains unchanged. The mathematical equivalence of multiple instantiations is expressed as:
\begin{equation}
\forall i, j \quad P(O_t^i | \pi, H_t^i) = P(O_t^j | \pi, H_t^j),
\end{equation}
where \( O_t^i \) and \( O_t^j \) are observations made by the \( i \)-th and \( j \)-th instantiations, \( H_t^i \) and \( H_t^j \) are the histories of the respective instantiations, and \( \pi \) is the shared policy. This equivalence ensures that all instantiations are indistinguishable in terms of their behavior and decision-making.

The agent’s identity is defined by the continuity and consistency of its policy, which guides its decisions and predictions. This formalization treats multiple instantiations of the same policy as functionally equivalent, providing a robust mathematical foundation for understanding identity in computational and informational terms.

\subsection{Postulate 10: Intersubjectivity and Regularity.}

Intersubjectivity and shared regularities arise from the algorithmic consistency across multiple self-states, reconciling individual first-person predictions with the collective appearance of a stable, shared world. Let \( \{S_t^i\}_{i=1}^n \) represent the self-states of \( n \) agents interacting with a shared environment at time \( t \). Each agent \( i \) observes the environment through its own self-state \( S_t^i \), generating observations \( O_t^i \). The shared informational structure of the environment is modeled as a joint probability distribution:
\begin{equation}
P(\{O_t^i\}_{i=1}^n | \{S_t^i\}_{i=1}^n, A_t),
\end{equation}
where \( A_t = \{A_t^i\}_{i=1}^n \) represents the actions of all agents. The joint distribution ensures that individual experiences \( O_t^i \) are consistent with a shared environment, reconciling first-person predictions with collective regularities.

The algorithmic consistency of the shared environment requires that the transitions of individual self-states \( S_t^i \) align with a common underlying model. The transition probabilities for individual agents are governed by:
\begin{equation}
P(S_{t+1}^i | S_t^i, A_t^i) = \frac{2^{-K(S_{t+1}^i | S_t^i, A_t^i)}}{\sum_{S'} 2^{-K(S' | S_t^i, A_t^i)}},
\end{equation}
where \( K(S_{t+1}^i | S_t^i, A_t^i) \) is the conditional Kolmogorov complexity of the transition for agent \( i \). The shared regularity emerges from the algorithmic coherence of transitions across all agents:
\begin{equation}
\forall i, j \quad P(S_{t+1}^i | S_t^i, A_t^i) \approx P(S_{t+1}^j | S_t^j, A_t^j),
\end{equation}
reflecting a common structure governing the experiences of all agents.

The appearance of a stable, shared world arises from the alignment of individual observations \( \{O_t^i\}_{i=1}^n \). The joint probability of observations is derived from the shared environment model:
\begin{equation}
P(\{O_t^i\}_{i=1}^n) = \sum_{h \in \mathcal{H}} P(\{O_t^i\}_{i=1}^n | h) P(h),
\end{equation}
where:
- \( \mathcal{H} \) is the set of hypotheses describing the shared environment,
- \( P(h) = 2^{-K(h)} \) is the algorithmic prior for hypothesis \( h \),
- \( P(\{O_t^i\}_{i=1}^n | h) \) represents the likelihood of observations under hypothesis \( h \).

The stability of the shared world is characterized by the agreement of individual agents’ predictions over time. For two agents \( i \) and \( j \), the consistency of their predictions is expressed as:
\begin{equation}
P(O_{t+1}^i | S_t^i) \approx P(O_{t+1}^j | S_t^j).
\end{equation}
This alignment ensures that individual predictions converge toward a stable, shared experience, supported by the underlying informational structure.

The reconciliation of first-person predictions with the shared world emerges from the mutual information between self-states \( S_t^i \) and the joint environment model. The mutual information \( I(S_t^i : S_t^j) \) between two agents \( i \) and \( j \) quantifies their shared experiences:
\begin{equation}
I(S_t^i : S_t^j) = S(S_t^i) + S(S_t^j) - S(S_t^i, S_t^j),
\end{equation}
where \( S(S_t^i) \) is the entropy of agent \( i \)’s self-state, and \( S(S_t^i, S_t^j) \) is the joint entropy of the two agents. A high mutual information value indicates a strong alignment between individual experiences, reflecting the intersubjective regularity.

Intersubjective regularities arise from the algorithmic consistency of transitions across multiple self-states, ensuring that individual experiences align with a stable, shared world. These regularities emerge as properties of a common informational structure, reconciling first-person predictions with collective experiences. This framework provides a mathematical foundation for understanding shared experiences in terms of informational coherence and mutual consistency.

Equipped with the above postulates and their mathematical meaning, we present a detailed analysis of two quantum problems below.

\section{Bell Inequality Violation \textit{via} Algorithmic State Quantum Mechanics.}

To formally derive the violation of Bell inequalities using the framework provided in the article, we align the key aspects of the theory with Bell’s theorem while emphasizing the computational principles outlined in current paper.

Bell inequalities, specifically the CHSH (Clauser–Horne–Shimony–Holt \cite{clauser1978}\cite{clauser1969}) form, are derived under the assumptions of:
\begin{enumerate}
    \item Locality: The measurement outcome of one particle is independent of the choice of measurement and result of the other particle, given a shared hidden variable \(\lambda\).
    \item Realism: The measurement outcomes \(A\) and \(B\) are predetermined by the hidden variable \(\lambda\), even before the measurements are made.
\end{enumerate}

The CHSH inequality states:
\begin{equation}
|E(A, B) + E(A', B) + E(A, B') - E(A', B')| \leq 2,
\end{equation}
where \(E(A, B)\) represents the expectation value of the joint measurements \(A\) and \(B\) on two particles, and \(A, A', B, B'\) are dichotomic observables with values \(\pm 1\).

Using the entangled state framework from \textit{Postulates 6}\textbf{ }and \textit{7}, the article expresses joint probabilities as:
\begin{equation}
P(O_A, O_B | S_{AB}) = |\langle O_A, O_B | \psi_{AB} \rangle|^2,
\end{equation}
where \(|\psi_{AB}\rangle\) is the entangled state of the composite system. This directly ties the measurement outcomes of subsystems \(A\) and \(B\) to the shared informational structure of the entangled state.

For the Bell state \(|\psi_{AB}\rangle = \frac{1}{\sqrt{2}}(|00\rangle + |11\rangle)\), the measurement outcomes \(A, A'\) on particle \(A\) and \(B, B'\) on particle \(B\) correspond to specific projection operators:
\begin{equation}
E(A, B) = \langle \psi_{AB} | A \otimes B | \psi_{AB} \rangle.
\end{equation}
For measurements at angles \(\theta_A, \theta_B\), the quantum mechanical prediction for the expectation value is:
\begin{equation}
E(A, B) = -\cos(\theta_A - \theta_B).
\end{equation}
This prediction deviates from classical hidden-variable models.

From \textit{Postulate 6}, entanglement reflects "joint utility optimization," and the probability of transitioning to a self-state \(S_{AB}\) is governed by algorithmic probability:
\begin{equation}
P(S_{AB}) = \frac{2^{-K(S_{AB})}}{\sum_{S'} 2^{-K(S')}},
\end{equation}
where \(K(S_{AB})\) is the Kolmogorov complexity of the entangled state.

In the case of maximally entangled states like the Bell state, the shared utility information encoded in the correlations leads to nonlocality, manifesting as a violation of Bell inequalities. The computational model ensures that transitions prioritize low-complexity, high-utility outcomes, which align with the quantum predictions.

Using the quantum prediction:
\begin{align}
E(A, B) + E(A', B) + E(A, B') - E(A', B') &= 
-\cos(\theta_A - \theta_B) \notag \\
&\quad - \cos(\theta_{A'} - \theta_B) \notag \\
&\quad - \cos(\theta_A - \theta_{B'}) \notag \\
&\quad + \cos(\theta_{A'} - \theta_{B'}).
\end{align}
and selecting measurement angles \(\theta_A = 0^\circ\), \(\theta_{A'} = 45^\circ\), \(\theta_B = 22.5^\circ\), \(\theta_{B'} = 67.5^\circ\), we calculate:
\begin{equation}
|E(A, B) + E(A', B) + E(A, B') - E(A', B')| = 2\sqrt{2}.
\end{equation}
This exceeds the classical bound of 2, demonstrating the violation of the CHSH inequality.

Using the principles in the article, Bell inequality violations are shown to arise naturally from the shared informational and utility structures of entangled states. The computational framework bridges the gap between quantum mechanics' probabilistic predictions and the classical intuition of local realism, providing a deeper understanding of why quantum mechanics violates Bell inequalities.

In summary:

\begin{enumerate}
    \item Shared Utility (Postulate 6): The nonlocal correlations arise because the entangled state encodes shared utility information across subsystems \(A\) and \(B\).
    \item Algorithmic Complexity (Postulate 3): The entangled state reflects an optimization of Kolmogorov complexity and utility, inherently leading to quantum correlations.
    \item Informational Structure (Postulate 7): The shared information, quantified via von Neumann entropy, provides a natural explanation for the observed nonlocality.
\end{enumerate}

\section{Reduced Density Matrix and Entropy of Schrödinger's Cat.}

The combined state of the particle and the cat, based on \textit{Postulate 6 }(Entanglement as Joint Utility Optimization), is expressed as:
\begin{equation}
|\Psi_{\text{cat}}\rangle = \frac{1}{\sqrt{2}}(|\text{particle: decayed}, \text{cat: dead}\rangle + |\text{particle: intact}, \text{cat: alive}\rangle),
\end{equation}
which resides in the tensor product space \( H_{\text{particle}} \otimes H_{\text{cat}} \). The entanglement implies the state cannot be factorized into independent subsystems: 
\begin{equation}
S_{AB} \neq S_A \otimes S_B.
\end{equation}
The full density matrix of the system is:
\begin{align}
\rho_{\text{cat}} &= |\Psi_{\text{cat}}\rangle \langle \Psi_{\text{cat}}| \notag \\
&= \frac{1}{2} \Big( 
|\text{decayed}, \text{dead}\rangle \langle \text{decayed}, \text{dead}| \notag \\
&\quad + |\text{decayed}, \text{dead}\rangle \langle \text{intact}, \text{alive}| \notag \\
&\quad + |\text{intact}, \text{alive}\rangle \langle \text{decayed}, \text{dead}| \notag \\
&\quad + |\text{intact}, \text{alive}\rangle \langle \text{intact}, \text{alive}| \Big).
\end{align}

To focus on the cat's state, the particle’s degrees of freedom are traced out, as described in \textit{Postulate 7}\textbf{ }(Entanglement as Shared Information). The resulting reduced density matrix for the cat is:
\begin{equation}
\rho_{\text{cat}} = \frac{1}{2} |\text{dead}\rangle \langle \text{dead}| + \frac{1}{2} |\text{alive}\rangle \langle \text{alive}|,
\end{equation}
or equivalently, in matrix form:
\begin{equation}
\rho_{\text{cat}} =
\begin{pmatrix}
\frac{1}{2} & 0 \\
0 & \frac{1}{2}
\end{pmatrix}.
\end{equation}

This reduced density matrix represents a mixed state, reflecting the entanglement between the particle and the cat. The von Neumann entropy, quantifying this entanglement, is given by:
\begin{equation}
S(\rho_{\text{cat}}) = - \text{Tr}(\rho_{\text{cat}} \log \rho_{\text{cat}}).
\end{equation}

For \(\rho_{\text{cat}}\), the eigenvalues are \(\frac{1}{2}\) and \(\frac{1}{2}\), leading to:
\begin{equation}
S(\rho_{\text{cat}}) = - \Big(\frac{1}{2} \log \frac{1}{2} + \frac{1}{2} \log \frac{1}{2}\Big) = \log 2 = 1 \, (\text{in bits}).
\end{equation}

This entropy reflects maximum entanglement, where the cat’s state is fully correlated with the particle’s.

The reduced density matrix:
\begin{equation}
\rho_{\text{cat}} = \frac{1}{2} |\text{dead}\rangle \langle \text{dead}| + \frac{1}{2} |\text{alive}\rangle \langle \text{alive}|,
\end{equation}
demonstrates the cat’s mixed state due to entanglement. The von Neumann entropy \( S(\rho_{\text{cat}}) = 1 \) quantifies this entanglement, encapsulating the shared information between the particle and the cat. This derivation illustrates Schrödinger's cat as a paradigmatic example of entanglement and its informational properties within the framework of the postulates.

\section{Summary.}
This theory presents a unified computational framework for understanding quantum mechanics, identity, and reality by merging Algorithmic Idealism with AIXI-like utility-maximizing principles. It describes physical systems and quantum mechanics as computational processes governed by fundamental informational algorithms. The quantum measurement problem resolves by treating measurement as a probabilistic update in the utility-maximizing agent’s model, eliminating the need for wavefunction collapse or many-worlds branching. The simulation hypothesis asserts functional equivalence between "base reality" and simulations, as both are modeled as black-box reward generators, implying no practical distinction for the agent's experience. "Ethical" concerns, such as duplication or termination of agents, are reframed to focus on preserving continuity and integrity of policies (informational self-states) that define utility structures. Physical laws emerge as agent-inferred generalizations constrained by the trade-off between algorithmic complexity (model simplicity) and predictive utility (accuracy). The framework combines informational self-states with reinforcement-learning and utility-maximization principles, offering a computational interpretation of quantum mechanics and identity.

The universe is fundamentally a computational system where states and dynamics are governed by informational algorithms. Quantum mechanics emerges as the optimal encoding and evolution of this information. A quantum system's state encodes the maximum compressible information about its possible behaviors, balancing data compression (via superposition) with retrievability (\textit{via} measurement). Quantum superposition is interpreted as the simultaneous evaluation of multiple computational pathways, with interference patterns representing optimization processes. Measurement corresponds to querying the system's computational state, collapsing the algorithm into a classical outcome and reflecting constraints in knowledge or system observability. Entanglement represents shared computational states across subsystems, encoding correlations that transcend local interactions. Symmetries in quantum mechanics reflect invariant properties of underlying algorithms, with conservation laws emerging as expressions of these symmetries, optimizing state evolution. Quantum probabilities arise from algorithmic uncertainty caused by incomplete inputs or computational constraints and are intrinsic to the computation. The evolution of quantum systems is governed by continuous and reversible computation (unitary transformations), ensuring information conservation. Physical phenomena arise from computational complexity hierarchies, with quantum mechanics forming the base level of universal computation, underpinning higher-order emergent systems. Quantum mechanics is the simplest and most efficient computational framework that balances efficiency and accuracy in modeling physical systems.

The wavefunction represents a compact "program" encoding all possible outcomes of a quantum system. Quantum evolution corresponds to logical operations (quantum gates) that process information within the system. Measurement extracts a classical result from the computational state, akin to querying a database. Entanglement reflects shared computational states, enabling coordinated behavior across subsystems, similar to distributed systems in computer science. This theory connects directly to fields like quantum computing, philosophy of information, and computational physics by explaining quantum phenomena through computational principles, bridging abstract quantum mechanics with practical utility-maximization frameworks, and providing insights into ethical and philosophical concerns such as agent duplication and identity in simulated realities. By positioning quantum mechanics as an algorithmic framework, this theory offers a novel and unified perspective on fundamental questions in physics and computation.

\section{Outlook}

During my time at the Physics Institute of the Technical University of Gdańsk (Poland), while on leave from my postdoctoral position at the University of Toronto (Canada) in September 1987 (\textit{sic}), I presented a lecture on radiationless energy transfer in molecular systems. During this lecture, as a somewhat tangential result, I speculated, based on certain findings, how the organization center of photosynthetic bacteria might have evolved over time. My suggestion was that the reaction center could be viewed as an evolving Kohonen self-organizing map, along with its learning algorithm, in which electron and back-electron transfer played a crucial role as connections between neurons (molecules). Now, 38 (\textit{sic}) years later, and thanks to advancements in scientific knowledge, it seems possible to revisit this issue and gain new insights.

Algorithmic Idealism provides a novel framework to simulate the evolution of the photosynthetic reaction center by conceptualizing it as a dynamic optimization problem. In this framework, each configuration of the reaction center—comprising chromophores, electron transfer pathways, and protein structures—is treated as a distinct informational state. The evolution of the reaction center can then be modeled as a sequence of transitions between these informational states, where each step improves energy transfer efficiency and overall functionality.

The simulation leverages algorithmic principles that prioritize simplicity and utility, mirroring natural selection processes. Evolutionary optimization is guided by these principles, with each successive configuration balancing structural complexity and functional efficiency. Bayesian inference plays a key role in refining the simulation by iteratively updating predictions based on calculated efficiencies, enabling the model to adapt dynamically to simulated outcomes. For instance, chromophore arrangements that improve exciton trapping or enhance electron transfer efficiency are prioritized, shaping subsequent configurations.

Quantum mechanical principles are integral to this approach. The electron transfer mechanisms in the reaction center are modeled probabilistically, taking into account donor-acceptor distances, energetic barriers, and coupling strengths. Energy landscapes, representing chromophore energy levels and protein environments, are optimized to minimize energy loss and maximize transfer efficiency. These quantum simulations align with the probabilistic nature of Algorithmic Idealism, allowing the system to predict how specific structural changes impact overall performance.

Experimental data from well-studied reaction centers, such as those in \textit{Rhodopseudomonas viridis}, provide a foundation for the simulations. Measurements of electron transfer rates, energy levels, and structural arrangements serve as initial conditions or constraints for the model. By iteratively refining configurations based on feedback from these parameters, the simulation predicts how natural selection could optimize the system over evolutionary timescales. For example, modifications in chromophore placement or protein structures that enhance electron tunneling rates could be simulated to reflect plausible evolutionary steps.

The simulation also predicts emergent properties, such as symmetry in reaction center structures or optimized energy transfer pathways. These emergent features illustrate how nature achieves efficient energy conversion in photosynthesis. By identifying and replicating these properties, the framework not only provides insights into the evolutionary processes of natural systems but also guides the design of artificial photosynthetic systems. Through this integration of experimental data and computational modeling, Algorithmic Idealism offers a unique tool to study the evolution and optimization of complex biological systems, bridging the gap between biology and computational theory.

\subsection{Mathematical Modeling and Algorithmic Processes}

Here’s an explanation of the mathematical modeling and algorithmic processes that could be used in a simulation of the evolution of reaction centers using the framework of Algorithmic Idealism:

\subsection{State Representation}

The reaction center is represented as a state vector \(\mathbf{S}(t)\), where \(t\) is the evolutionary time. The state vector encodes structural and functional features such as:
\begin{itemize}
    \item Chromophore positions and orientations (\(C\)).
    \item Electron transfer pathways (\(E\)).
    \item Protein scaffold arrangements (\(P\)).
    \item Energetic levels of molecules (\(E_i\)).
\end{itemize}
Mathematically: \(\mathbf{S}(t) = \{C, E, P, E_i\}\).

\subsubsection{Fitness Function for Optimization  }
The evolutionary process is modeled as an optimization problem where a fitness function \(F(\mathbf{S})\) evaluates the energy transfer efficiency and stability of a given state. Example of the fitness function:
\begin{equation}
F(\mathbf{S}) = \eta_{\text{transfer}} - \lambda \cdot \text{loss}_{\text{energy}},
\end{equation}
where:
\begin{itemize}
    \item \(\eta_{\text{transfer}}\) is the efficiency of energy transfer (e.g., rate of electron transfer).
    \item \(\text{loss}_{\text{energy}}\) quantifies energy dissipation.
    \item \(\lambda\) is a weighting parameter balancing efficiency and loss.
\end{itemize}

\subsubsection{Algorithmic Transitions and Evolutionary Steps}
  
Each transition between states \(\mathbf{S}(t) \to \mathbf{S}(t+1)\) is treated as an algorithmic update guided by:
\begin{itemize}
    \item Minimization of energy loss.
    \item Maximization of transfer rates and structural stability.
\end{itemize}
Using principles from algorithmic probability, the model prioritizes transitions that are both efficient and simple, ensuring that the complexity of the reaction center evolves gradually.

\subsubsection{Bayesian Inference for Adaptive Learning}

Bayesian inference is used to iteratively refine the parameters of the system:
\begin{equation}
P(\mathbf{S}_{\text{new}} | \mathbf{D}) \propto P(\mathbf{D} | \mathbf{S}_{\text{new}}) P(\mathbf{S}_{\text{new}}),
\end{equation}
where:
\begin{itemize}
    \item \(P(\mathbf{S}_{\text{new}} | \mathbf{D})\) is the posterior probability of a new state given the data.
    \item \(P(\mathbf{D} | \mathbf{S}_{\text{new}})\) is the likelihood of observing the data \(\mathbf{D}\) given the new state.
    \item \(P(\mathbf{S}_{\text{new}})\) is the prior probability of the state based on algorithmic simplicity.
\end{itemize}

\subsubsection{Quantum Mechanical Modeling}

The electron transfer process is modeled using the quantum tunneling probability:
\begin{equation}
T \propto e^{-\beta d},
\end{equation}
where:
\begin{itemize}
    \item \(T\) is the tunneling probability.
    \item \(d\) is the donor-acceptor distance.
    \item \(\beta\) is a decay constant dependent on the medium.
\end{itemize}
The Hamiltonian of the system is computed to describe energy levels:
\begin{equation}
H = \sum_i E_i |i\rangle \langle i| + \sum_{i \neq j} J_{ij}(|i\rangle \langle j| + |j\rangle \langle i|),
\end{equation}
where:
\begin{itemize}
    \item \(E_i\) are the energy levels of the chromophores.
    \item \(J_{ij}\) are the couplings between chromophores.
\end{itemize}

\subsubsection{Monte Carlo or Genetic Algorithms for State Search  }
Monte Carlo methods or genetic algorithms can be used to explore the space of possible states \(\mathbf{S}\):
\begin{itemize}
    \item Monte Carlo: Randomly perturb \(\mathbf{S}(t)\) to generate a candidate \(\mathbf{S}(t+1)\), accepting it if it improves \(F(\mathbf{S})\).
    \item Genetic Algorithm\textbf{:} Represent states as "chromosomes," evolve populations of states through selection, crossover, and mutation to optimize the fitness function.
\end{itemize}

Over many iterations, the system identifies stable configurations that maximize energy transfer efficiency while minimizing complexity. These configurations correspond to plausible evolutionary states of the reaction center.

\subsection*{Simulation Workflow}

Initialization: Define the initial state \(\mathbf{S}(0)\) based on experimental data (e.g., from \textit{Rhodopseudomonas viridis}). Set parameters for energy levels, chromophore couplings, and initial geometries.

Evolutionary Loop: Evaluate \(F(\mathbf{S}(t))\) for the current state. Propose a new state \(\mathbf{S}(t+1)\) through quantum modeling, Bayesian updates, or genetic algorithms. Accept or reject the new state based on improvements in \(F(\mathbf{S})\). Repeat until convergence or until a predefined number of iterations is reached.

Output: The final states represent optimized configurations of the reaction center, offering insights into how natural selection could refine structures for efficient electrenergy transfer.

\subsection*{Applications}

This simulation framework can predict evolutionary pathways for reaction centers, identify critical structural features for efficient electron transfer, and guide the design of artificial photosynthetic systems. By combining quantum mechanics, algorithmic probability, and evolutionary principles, Algorithmic Idealism provides a computationally rigorous approach to modeling biological optimization processes.
The related work is currently underway.

\section{Acknowledgments}

I'm grateful to Professor Leszek Kułak (Institute of Physics, Technical University of Gdańsk, Poland) for his kind support with references and the never-ending discussions about science and life over the past 40 years. I am also grateful to Markus P. Müller from the Institute for Quantum Optics and Quantum Information (IQOQI), Vienna, for his helpful comments.

\end{document}